\documentclass[manuscript, screen, nonacm]{acmart}

\AtBeginDocument{%
  }

\setcopyright{none}
\acmConference[CHI '25]{Conference on Human Factors in Computing Systems}{April 26--May 1, 2025}{Yokohama, Japan}

\begin{document}
\title{Aspirational Identities: How VTubing Affords New Modes of Digital Embodiment}

\author{Wells Lucas Santo}
\email{ephemera@umich.edu}
\orcid{0009-0000-0748-0105}
\affiliation{%
  \institution{University of Michigan}
  \city{Ann Arbor}
  \state{Michigan}
  \country{USA}
}

\author{Hibby Thach}
\email{hibby@umich.edu}
\orcid{0000-0002-0520-0218}
\affiliation{%
  \institution{University of Michigan}
  \city{Ann Arbor}
  \state{Michigan}
  \country{USA}
}

\ccsdesc[500]{Human-centered computing~Interaction paradigms}

\keywords{digital embodiment, trans technologies, digital yellowface, VTubers}

\begin{abstract}
VTubing, short for "Virtual YouTubing", is a modality of streaming where a CGI avatar is used in place of the streamer's analog face and body. As VTubing has become increasingly popular, especially for queer and trans streamers of color, it is worth studying as a way for streamers to express identities that they hope to embody, which we call "aspirational identities". In this workshop submission, we draw on preliminary findings from qualitative research to argue that VTubing is a potent object of study for understanding digital embodiment and the performance of identity online, especially for questions of race, gender, sexuality, and disability.
\end{abstract}

\maketitle

\section{VTubing and Digital Embodiment}

On October 10, 2024, disabled Puerto Rican streamer Ironmouse broke the record for having the most subscribers on Twitch, the largest live-streaming platform globally, receiving over 326,000 paid subscriptions\footnote{Subscriptions on Twitch are tiered, with higher tier subscribers gaining additional content from the channel they are subscribed to. The lowest tier cost for an Ironmouse subscription is \$5.99/month, which given 326,000 subscribers, translates to a total revenue of roughly \$1.6M/month. Twitch Partners such as Ironmouse receive a 70\% revenue share, meaning that from these subscriptions, she received at minimum \$1.1M. Twitch users do not have to renew their monthly subscriptions, and Ironmouse typically averages 15,000 subscribers each month.} \cite{christophercruzIronmouse2024}. What sets Ironmouse apart from most streamers, though, is the fact that she is a VTuber, a particular type of streamer who performs using a CGI avatar instead of their own analog face and body on camera. In an interview with the podcast Press ESC, Ironmouse discussed her decision to stream as a VTuber, explaining, "I realized very quickly that I did not want to be on camera. I have medical devices on me, I’m on oxygen, I have a port in my chest... yeah, I’m like a cyborg!" \cite{valkyraeIronMouses} Though Ironmouse is not explicitly invoking the concept of \textit{cyborgs} from Donna Haraway \cite{harawayManifesto1985}, VTubing has allowed her to act with and \textit{enact} a body multiple \cite{molBody2002} --- an amalgamation of herself and "non-human" objects, such as computer hardware, motion capture software, rigging software, and a digital avatar --- that together constitute her VTuber identity. She is both digital \textit{and} analog, existing in parallel, as multiple.

\begin{figure}[h]
  \centering
  \includegraphics[width=0.5\hsize]{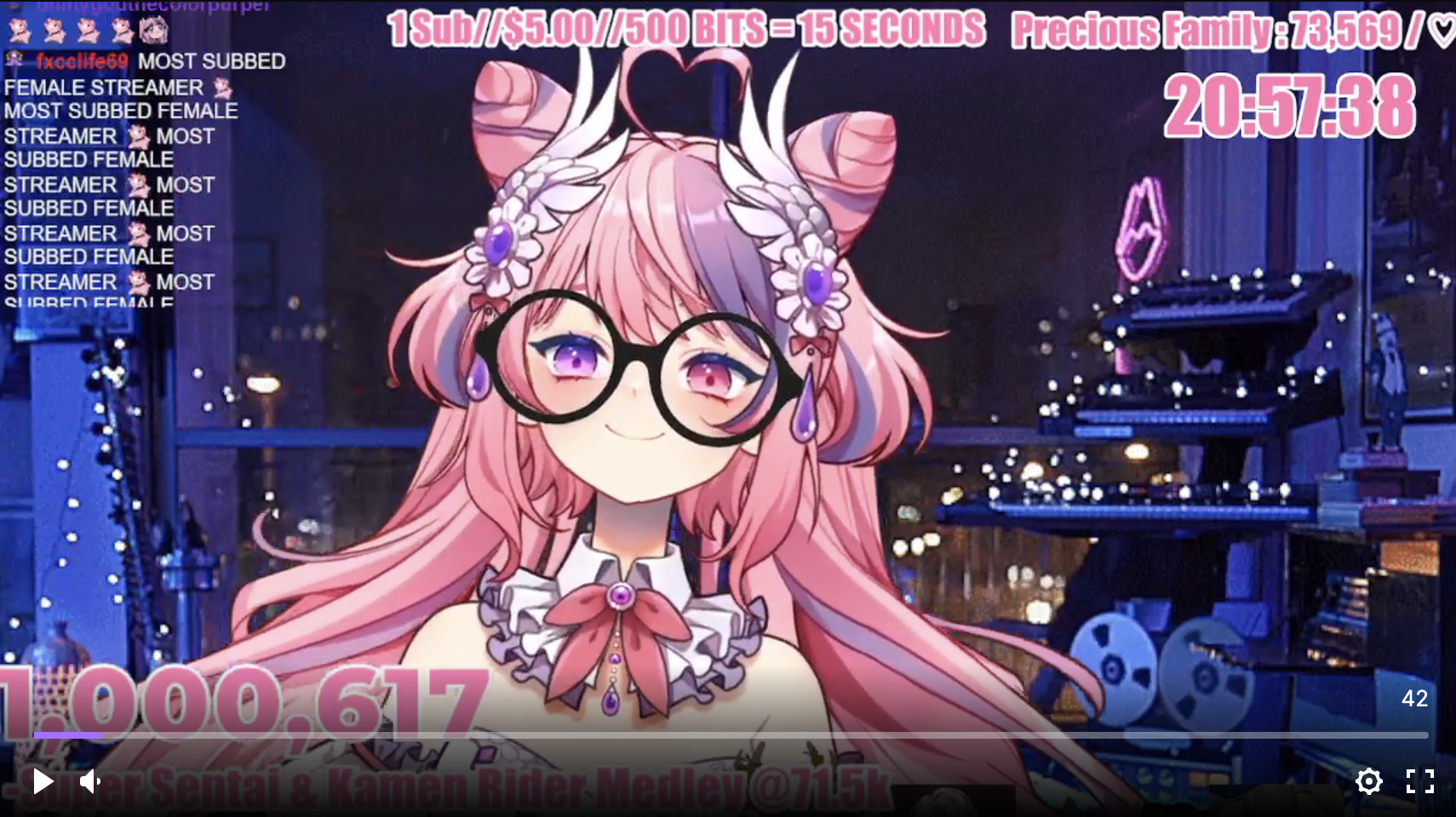}
  \caption{Screenshot of an Ironmouse Twitch clip.}
  \Description{Ironmouse}
  \label{fig:ironmouse}
\end{figure}

VTubing is distinct from previous forms of digital avatars in that it is operated not by a series of keypresses or mouseclicks, but rather by the movement of the face and body, tracked using computer vision algorithms. Like with the use of a motion capture suit, or various sensors for full-body tracking in VR, VTubing technologies allow streamers to \textit{directly} interface with their digital bodies, allowing for a more immersive sense of embodiment. However, because it does not need the typically expensive equipment required for traditional motion-capture, VTubing has grown to become an international phenomenon that provides users around the world with a more affordable interaction paradigm, one that can capture granular movements and facial expressions in real-time. When you smile in the analog world, your VTuber avatar smiles with you. When you blink, they blink. The digital body is \textit{felt} as an extension of one's self, which we explore in our ethnographic and autoethnographic work.

\section{Ongoing Work and Methods}

\subsection{Interviews and Ethnography}

Both authors have collected empirical qualitative data about VTubers for ongoing studies, each addressing different but related research questions. In order to analyze the embodied experiences of multi-marginalized\footnote{We define multi-marginalized as identifying as marginalized across more than one axis of identity (gender, sexuality, race, disability, etc.).} streamers more broadly, the second author led an interview study of multi-marginalized streamers on Twitch ($N$ = 15). From these interviews, in which seven streamers used VTubing technologies, we found that VTubing allowed streamers the opportunity to digitally embody their aspirational selves or express aspects of their identities that they felt could not be expressed in their daily lives. For example, on the topic of performing drag, the trans and gender non-conforming VTuber Ezra shared, "I feel like [VTubing] is the easiest way to [perform drag] because I would love to do drag like IRL, but logistically, no, it’s never going to happen." From pilot participant-observation work, the first author has engaged with queer, Asian American VTubers at OffKai Expo, the premiere VTuber convention, and online via Discord, an instant messaging platform where VTubers often chat with their fans and community members. In these settings, VTubers have expressed how VTubing has been both liberatory and playful. The non-binary streamer Ying, for example, who is often misgendered as an Asian American woman in real-life, specifically commissioned a VTuber model that looked androgynous, with a balance of masculine and feminine physical traits, which allowed them perform on stream while being seen in the way they wanted. Similarly, the artist Fuwa, an Asian American woman, began to use a frog VTuber model so that she could still express herself through body movements and facial expressions on-screen, but without having to deal with a feminized body perceived through the lens of the male gaze.

\begin{figure}[h]
  \centering
  \includegraphics[width=0.3\hsize]{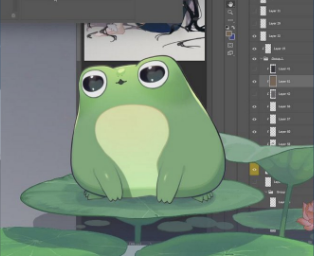}
  \caption{Screenshot of a Fuwa Twitch clip where she is using her frog VTuber model.}
  \Description{Fuwa}
  \label{fig:fuwa}
\end{figure}

\subsection{Auto-Ethnography}

Both authors, who are trans people of color, have used VTubing technologies on the Twitch streaming platform. The first author initially streamed as a woman-presenting VTuber in 2020, four years before starting feminizing hormone therapy. VTubing provided them a feeling of gender euphoria with being able to operate a (digital) body in real-time that aligned with the gender identity they desired. The second author began streaming in the year 2023 alongside beginning to wear makeup and feminine clothing. Employing collaborative autoethnographic \cite{changCollaborative2013} and critical autoethnographic \cite{boylornCritical2021} methods, the second author analyzed her own individual account as a queer transfemme of mixed Vietnamese heritage alongside the accounts of her interviewees, considering how both her and her interviewees' narratives speak to broader issues for multi-marginalized creators within toxic technocultures, or anti-progressivist platform cultures \cite{massanariGamergate2017}. Through streaming and eventually VTubing, she has been able to document her transition and play with gender in ways she would not be able to physically. For both authors, this auto-ethnographic work is ongoing.

\begin{figure}[h]
  \centering
  \includegraphics[width=0.3\hsize]{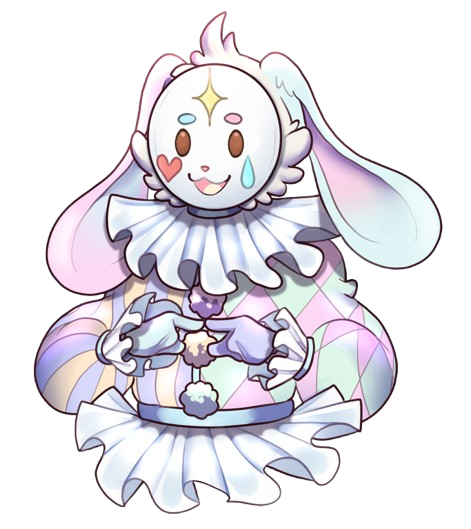}
  \caption{Hoppy, the VTuber model that the first author currently streams with. This model was designed by artist \href{https://www.remidavis.com/)}{Remi Davis.}}
  \Description{Hoppy}
  \label{fig:hoppy}
\end{figure}

\section{Identity Formation and Performance}

Based on ongoing empirical findings, we find that VTubing provides users the ability to explore aspirational and imaginative bodies, even those which are non-human. From robots to Lovecraftian monsters to frogs, there is no limit to the appearance of the digital body of the VTuber. In this way, VTubing allows for play with one's body and one's identity, bringing early Internet concepts of identity play on the textual web into our contemporary multimedia, live-streaming web. Just as Sherry Turkle \cite{turkleLife1995} once theorized, we are able to experiment with a multiple and fluid "self" in the digital. In the realm of the "human," this allows us to explore different embodied identities, across race, gender, and disability, which holds implications not only for the individual streamer, but for sociocultural formations of identity as well. Digital or not, bodies are deeply political, carrying historical and sociocultural meanings that map onto intersecting axes of identity. Such is true with the construction of the face, which can be seen as a technology that "does not come ready made, but is produced by an 'abstract machine of \textit{faciality},'" as Gilles Delueze and Félix Guattari have theorized \cite{deleuzeThousand1987}. Or more simply, as described by cultural theorist Claudio Celis Bueno, "faces have to be made and not all societies require making faces" \cite{celisbuenoFace2020}. It is in this sense that the face --- and by extension, the body --- operates as an assemblage, imbued and interpellated with meaning via social processes. Because VTubing requires a digital artist to design the virtual body and a streamer to perform with that body, we can interrogate VTubers in order to reveal the racial, gendered, and queer significations of our body. In other words, we can explore various intersecting identities (e.g., race, gender, and body type) using VTubing, which can serve as an insightful and rich method for interrogating body politics.

\subsection{Trans Embodiment}

One particularly relevant axis of identity for exploring digital embodiment through VTubing technologies is transness. The Internet has changed the experience of transness and transition for many individuals, with artifacts such as video game avatars allowing for the exploration, development, and rehearsal of aspirational gender identities \cite{morganRole2020, mckennaYou2024}. Bredikhina \cite{liudmilabredikhinaIdentity2020} finds that unlike these video game avatars, VTubers and PngTubers are specifically linked to one's embodied activity as a creator or entertainer. Lu et al. \cite{luMore2021} finds that Vtubers and PngTubers allow for varied performances of gender as well, as the range of activities and representations of gender are much less confined than with a video game avatar. Based on our initial findings, we are able to corroborate the above claims, while posing VTubing as a "trans technology," a term coined by Oliver Haimson, where trans streamers are able to utilize VTubing to help meet their trans-related needs, including embracing materiality, multiplicity, fluidity, and ambiguity \cite{haimsonDesigning2020a, starksDesigning2019, haimsonHow2023}.

\subsection{Digital Yellowface and Racial Formation}

However, VTubing has also raised questions of racial appropriation due to the common incorporation of Japanese, anime, or "kawaii" aesthetics into VTuber models, even though many English-speaking streamers are White. Members of the VTubing community along with the second author called this "digital yellowface," and have argued that there is a sort of "Asianness as default" that is being performed by these White streamers, who might also play up certain Asian stereotypes and incorporate fragmented Japanese phrases into their performance. Drawing from scholars such as Lisa Nakamura \cite{nakamuraRace1995} and Leslie Bow \cite{bow_racist_2022}, this can be conceptualized both as a form of identity tourism and Orientalist fetishization, which brings up questions of how race is enacted through the body. To go further, the second author argues that these VTuber performances re-contribute to the racial formation \cite{omiRacial2015} of "Asian" identities, driven partly by these White creators and their perceptions of Asianness as expressed in the body.


\section{Conclusion}

As VTubing has become an increasingly popular modality for online streamers, especially queer and trans streamers of color, we should no longer consider it a niche digital practice. Through interviews, ethnography, and autoethnography, we have found how VTubing technology allows for new ways to express aspirational identities and the co-construct one's embodied identity. However, we argue that VTubing is not a solution to issues of racism, sexism, queerphobia, and ableism in the digital; rather, it only serves as a potent object of study for understanding the sociopolitical nature of the body and what it signifies in terms of race, gender, sexuality, and disability. But despite the prominence and cultural impact of VTubing today, VTubers remain understudied in academic scholarship, whether by those studying streamer cultures or human-computer interfaces. Therefore, our work is motivated by a desire to discuss VTubing at the Body Politics workshop as a method, mode, and object of study for digital embodiment, the plurality of the body, and the realization that we are able to be "more-than-human" \cite{homewoodTracing2021}.

\bibliographystyle{ACM-Reference-Format}
\bibliography{BodyPolitics}

\end{document}